\documentclass[preprint]{aastex}

\shorttitle{Close binary stars.~IX}
\shortauthors{Pych \& et~al.}

\begin{document}

\title{Radial Velocity Studies of Close Binary
Stars.~IX\footnote{Based on the data obtained at the David Dunlap
Observatory, University of Toronto.}}

\author{Wojtek~Pych\altaffilmark{2},
Slavek~M.~Rucinski, Heide~DeBond, J.~R.~Thomson,
Christopher~C.~Capobianco, R.~Melvin~Blake}
\affil{David Dunlap Observatory, University of Toronto \\
P.O.~Box 360, Richmond Hill, Ontario, Canada L4C 4Y6}
\email{(pych,rucinski,debond,jthomson,capobianco,blake)@astro.utoronto.ca}
\and
\author{Waldemar~Og{\l}oza}
\affil{Mt. Suhora Observatory of the Pedagogical University \\
ul.~Podchor{\c{a}}{\.z}ych 2, 30-084 Cracow, Poland}
\email{ogloza@ap.krakow.pl}
\and
\author{Greg~Stachowski}
\affil{Copernicus Astronomical Center, Bartycka 18, 00--716
Warszawa, Poland}
\email{gss@camk.edu.pl}
\and
\author{Piotr Rogoziecki, Piotr Ligeza}
\affil{Adam Mickiewicz University Observatory, S{\l}oneczna 36,
60--286 Pozna\'{n}, Poland}
\email{progoz@moon.astro.amu.edu.pl, piotrl@amu.edu.pl}
\and
\author{Kosmas~Gazeas}
\affil{Department of Astrophysics, Astronomy and Mechanics\\
National \& Kapodistrian University of Athens,
GR~157~84 Zographou, Athens, Greece}
\email{kgaze@skiathos.physics.auth.gr}

\altaffiltext{2}
{On leave from: Copernicus Astronomical Center,
Bartycka 18, 00--716 Warszawa, Poland}

\begin{abstract}

Radial-velocity measurements and sine-curve fits to the orbital
velocity variations are presented for the eighth set of ten
close binary systems: AB~And, V402~Aur, V445~Cep, V2082~Cyg,
BX~Dra, V918~Her, V502~Oph, V1363~Ori, KP~Peg, V335~Peg.
Half of the systems (V445~Cep, V2082~Cyg, V918~Her, V1363~Ori,
V335~Peg) were discovered photometrically
by the {\it Hipparcos} mission and
all systems are double-lined (SB2) contact binaries.
The broadening function method permitted improvement of the orbital
elements for AB~And and V502~Oph. The other systems have been observed
for radial velocity variations for the first time; in this group are
five bright ($V<7.5$) binaries: V445~Cep, V2082~Cyg, V918~Her,
KP~Peg and V335~Peg.
Several of the studied systems are prime candidates for combined
light and radial-velocity synthesis solutions.
\end{abstract}


\section{INTRODUCTION}
\label{sec1}


\begin{deluxetable}{lccrrrccc}
\tabletypesize{\scriptsize}

\pagestyle{empty}
\tablecolumns{9}

\tablewidth{0pt}
\tablenum{2}
\tablecaption{Spectroscopic orbital elements}
\tablehead{
   \colhead{Name} &                
   \colhead{Type} &                
   \colhead{Other names} &         
   \colhead{$V_0$} &               
   \colhead{K$_1$} &               
   \colhead{$\epsilon_1$} &        
   \colhead{T$_0$ -- 2,400,000} &  
   \colhead{P (days)} &            
   \colhead{$q$}          \\       
   \colhead{}     &                
   \colhead{Sp.~type}    &         
   \colhead{}      &               
   \colhead{} &                    
   \colhead{K$_2$} &               
   \colhead{$\epsilon_2$} &        
   \colhead{(O--C)(d)~[E]} &         
   \colhead{$(M_1+M_2) \sin^3i$} &    
   \colhead{}                      
}
\startdata

AB And & EW/W & SAO~73069 & $-27.53(0.67)$ & $130.32(1.17)$ & 5.13 & 52,503.0443(4)  & $0.3318919$ & $0.560(7)$ \\
       & G8V & HIP~114508 &          & $232.88(0.83)$ & 8.20 & $-0.0105$~$[-2957]$ & 1.648(20) &          \\[1mm]

V402 Aur & EW/W & HD~282719 & $+40.82(0.93)$ & $41.94(0.83)$ & 4.33 & 52,448.9698(16) & $0.603491$ & $0.201(6)$ \\
    & F2V & HIP~23433 &          & $208.98(2.17)$ & 13.46 & $+0.0008$~$[371]$ & 0.988(27) &          \\[1mm]

V445 Cep & EW/A & HD~210431 & $+40.69(0.95)$ & $20.33(0.85)$ & 4.57 & 52,470.5847(21) & $0.448776$ & $0.167(10)$ \\
    & A2V & HIP~109191 &          & $122.08(2.04)$ & 11.55 & $+0.0072$~$[8847]$ & 0.134(6) &          \\[1mm]
 
V2082 Cyg & EW/A: & HD~183752 & $-34.12(0.58)$ & $33.16(0.51)$ & 2.85 & 52,466.1122(17) & $0.714084$ & $0.238(5)$ \\
    & F2V & HIP 95833 &          & $139.38(0.99)$ & 6.75 & $-0.0249$~$[5554]$ & 0.380(7) &          \\[1mm]
 
BX Dra & EW/A & HIP~78891 & $-26.11(3.43)$ & $80.01(2.46)$ & 8.87 & 52,248.2984(34) & $0.579027$ & $0.289(16)$ \\
    & F0IV-V & &          & $276.39(6.37)$ & 23.90 & $+0.0006$~$[6473]$ & 2.72(16) &          \\[1mm]
 
V918 Her & EW/W & HD~151701 & $-25.72(0.74)$ & $53.93(0.49)$ & 3.45 & 52,555.8419(14) & $0.57481$ & $0.271(5)$ \\
    & A7V & HIP~82253 &          & $199.37(1.72)$ & 10.43 & $+0.0253$~$[7055]$ & 0.968(21) &          \\[1mm]
 
V502 Oph & EW/W & HD~150484 & $-42.56(0.85)$ & $82.71(1.03)$ & 6.38 & 52,452.7492(7) & $0.45339$ & $0.335(9)$ \\
    & G0V & HIP~81703 &          & $246.70(1.00)$ & 6.89 & $-0.0500$~$[8718]$ & 1.679(22) &          \\[1mm]
 
V1363 Ori & EW/A & HD~289949 & $+37.89(2.02)$ & $44.88(2.46)$ & 16.32 & 52,592.4999(20) & $0.431921$ & $0.205(15)$ \\
    & F:\tablenotemark{a} & HIP~23809 &          & $219.30(4.18)$ & 21.62 & $+0.0196$~$[9475]$ & 0.825(45) &          \\[1mm]
 
KP Peg & EW/A & HD~204215 & $+5.52(1.03)$ & $72.84(1.05)$ & 6.20 & 52,504.9303(21) & $0.727203$ & $0.322(10)$ \\
    & A2V & HIP~105882 &          & $226.43(4.10)$ & 18.18 & $+0.0214$~$[5507]$ & 2.020(86) &          \\[1mm]
 
V335 Peg & EW/A: & HD~216417 & $-15.41(0.43)$ & $44.61(0.27)$ & 2.47 & 52,330.1642(15) & $0.810720$ & $0.262(4)$ \\
    & F5V & HIP~112960 &          & $170.56(1.77)$ & 12.54 & $-0.1590$~$[4724]$ & 0.837(21) &          \\[1mm]

\tablenotetext{a}{V1363 Ori: Early to mid F-type}

\tablecomments{The spectral types given in column two
are all new and relate to the combined spectral
type of all components in a system. The convention of naming the
binary components in this table is that the more massive
star is marked by the subscript ``1'', so that 
the mass ratio is defined to be always $q \le 1$. Figures 1 -- 3
should help identify which component is eclipsed at the primary
minimum. The standard errors of the circular 
solutions in the table are expressed in units of last decimal places 
quoted; they are given in parantheses after each value. 
The center-of-mass velocities ($V_0$), the 
velocity amplitudes (K$_i$) and the standard unit-weight 
errors of the solutions ($\epsilon$) are all expressed 
in km~s$^{-1}$. The spectroscopically determined 
moments of primary minima are given by $T_0$; the corresponding
$(O-C)$ deviations (in days) have been calculated from the most 
recent available ephemerides, as given in the text,
using the assumed periods and the number of epochs given by [E].
The values of $(M_1+M_2)\sin^3i$ are in the solar mass units.
}
\enddata
\end{deluxetable}

This paper is a continuation in a series of papers of
radial-velocity studies of close binary stars
\citep{ddo1,ddo2,ddo3,ddo4,ddo5,ddo6,ddo8} and presents
data for the eighth group of ten close binary stars observed
at the David Dunlap Observatory.
Selection of the targets is quasi-random: At a given time,
we observe a few dozen close binary systems with periods
shorter than one day, brighter than 11 magnitude and with
declinations $>-15^\circ$. We
publish the results in groups of ten systems as soon as reasonable
orbital elements are obtained from measurements evenly distributed
in orbital phases. For technical details and conventions,
and for preliminary estimates of errors and uncertainties,
see the interim summary paper \citet[hereafter Paper VII]{ddo7}.
With this paper, we decided to introduce some
minor changes into the reduction process:
We used the pair of IRAF routines
{\it noao.imred.spec.fitcoords} and {\it noao.imred.spec.transform} to
rectify images of the spectra and improve wavelength calibrations;
the procedure of cosmic ray removal was done using a separate,
standalone program \citep{Pych2003}.

We estimate spectral types of the program stars
using our classification spectra. These are compared with the
mean $(B-V)$ color indexes taken from the {\it Tycho-2\/}
catalog \citep{Tycho2}
and the photometric estimates of the spectral types
using the relations published by \citet{Bessell1979}.

The observations reported in this paper have been
collected mostly during the year 2002; exceptions
are: BX~Dra and V335~Peg, for which some observations
were collected in 2001, and V918~Her, for which some
observations were in May 2003. The ranges of dates for individual
systems can be found in Table~\ref{tab1}.

All systems discussed in this paper, except AB~And and V502~Oph,
have been observed for radial-velocity variations for the first
time.
We have derived the radial velocities in the same way as described
in previous papers. See Paper~VII for a discussion of the
broadening-function approach used in the derivation of
the radial-velocity orbit parameters:
the amplitudes, $K_i$, the center-of-mass
velocity, $V_0$, and the time-of-primary-eclipse epoch, $T_0$.

\begin{figure}
\figurenum{1}
\plotone{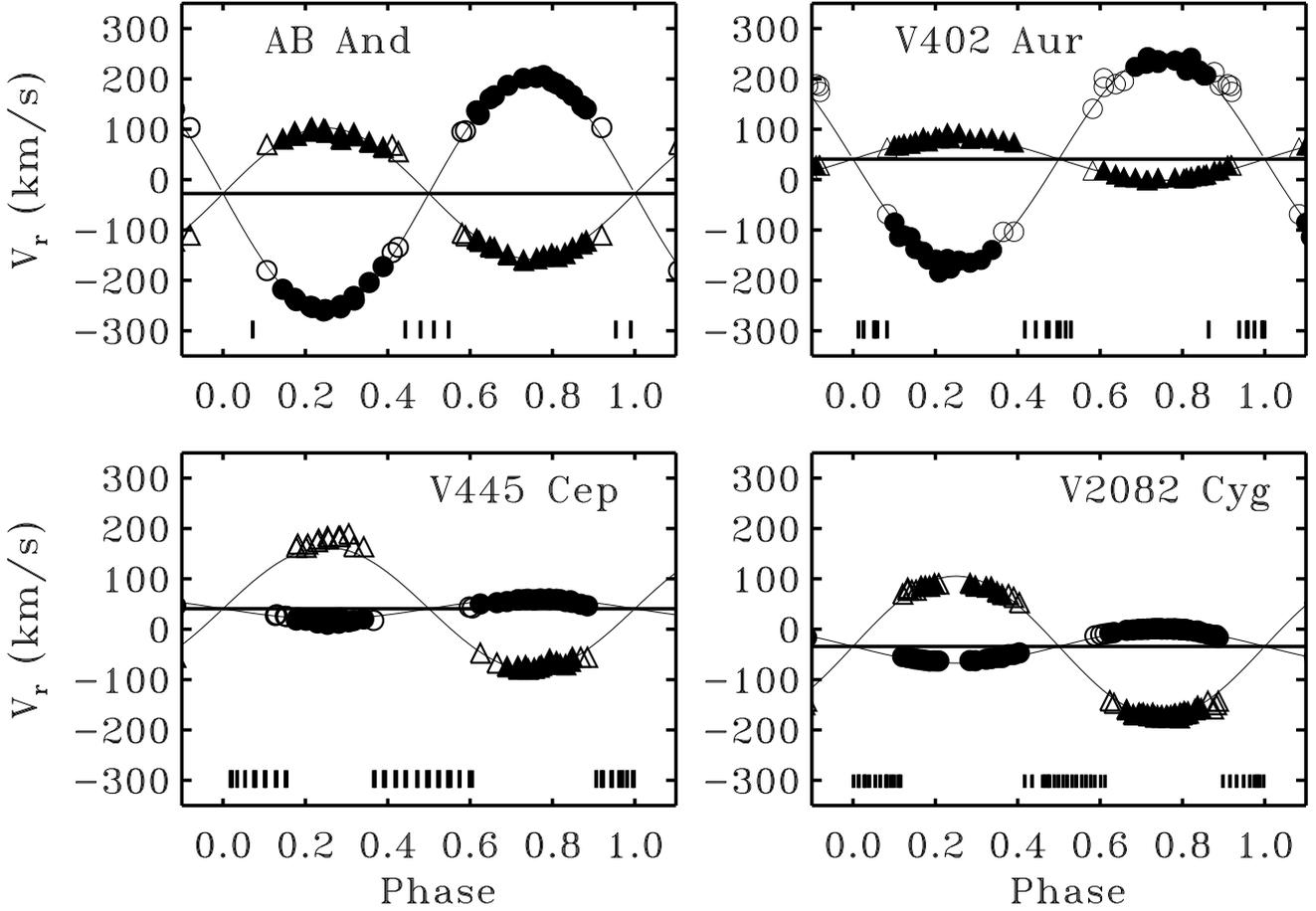}
\caption{
Radial velocities of the systems AB~And, V402~Aur, V445~Cep and
V2082~Cyg are plotted in individual panels versus the orbital phases.
All four systems are contact binaries.
V445~Cep and V2082~Cyg are A-type contact systems,
while AB~And and V402~Aur are W-type contact systems.
The lines give the respective circular-orbit (sine-curve)
fits to the radial velocities.
The circles and triangles in this and the next two figures
correspond to components with velocities $V_1$ and $V_2$,
as listed in Table~\ref{tab1}, respectively.
The component eclipsed at the minimum corresponding to
$T_0$ (as given in Table~\ref{tab2}) is the one which shows
negative velocities for the phase interval $0.0 - 0.5$.
The open symbols indicate observations contributing
half-weight data in the solutions. Short marks in the lower
parts of the panels show phases of available
observations which were not used in the solutions because of the
blending of lines. All panels have the same vertical
range, $-350$ to $+350$ km~s$^{-1}$.
}
\label{fig1}
\end{figure}

\begin{figure}
\figurenum{2}
\plotone{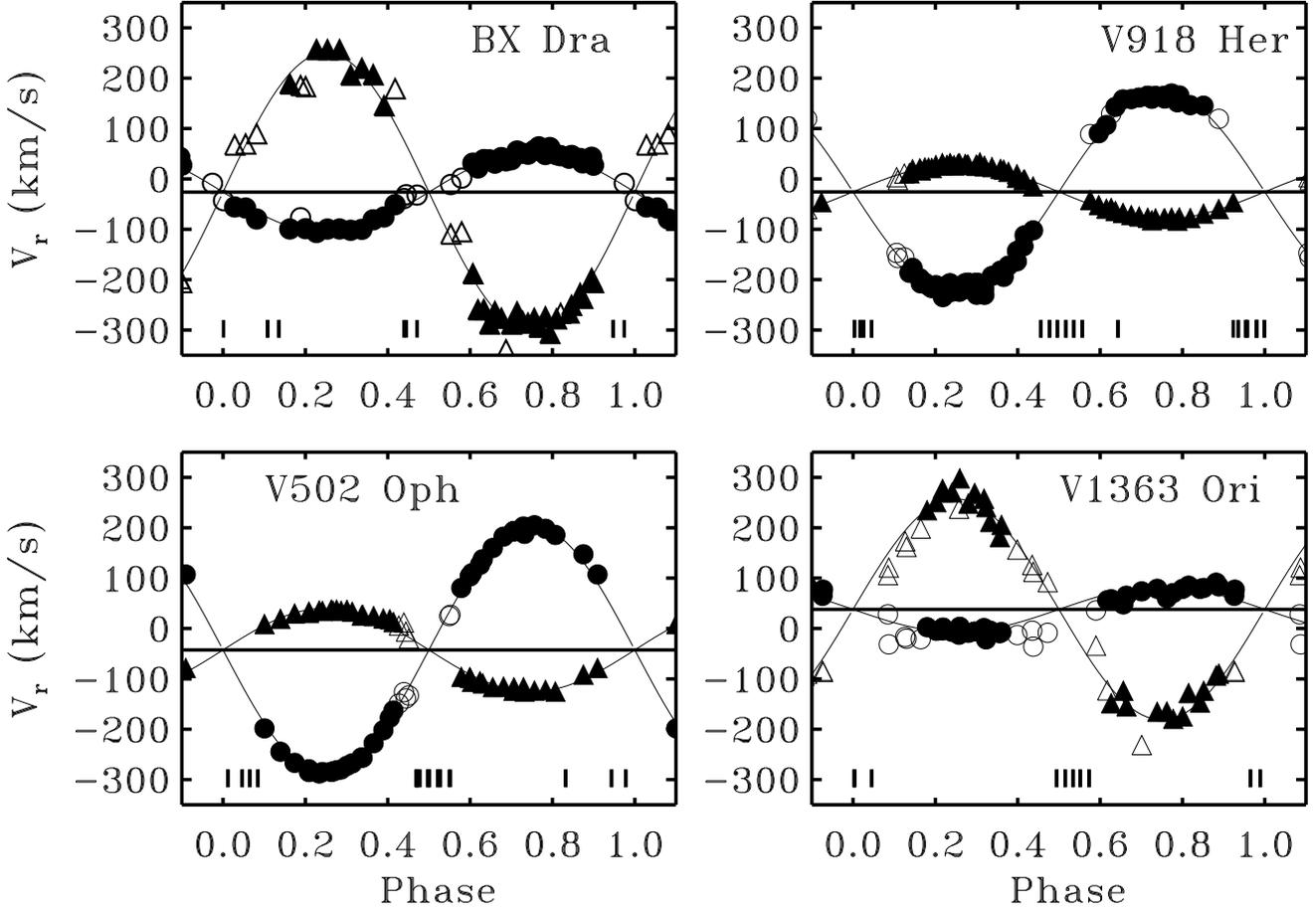}
\caption{
Same as for Figure~1, but with the radial velocity orbits
for the systems BX~Dra, V918~Her, V502~Oph and
V1363~Ori.
BX~Dra and V1363~Ori are A-type contact systems,
while V918~Her and V502~Oph are W-type contact systems.
}
\label{fig2}
\end{figure}

\begin{figure}
\figurenum{3}
\plotone{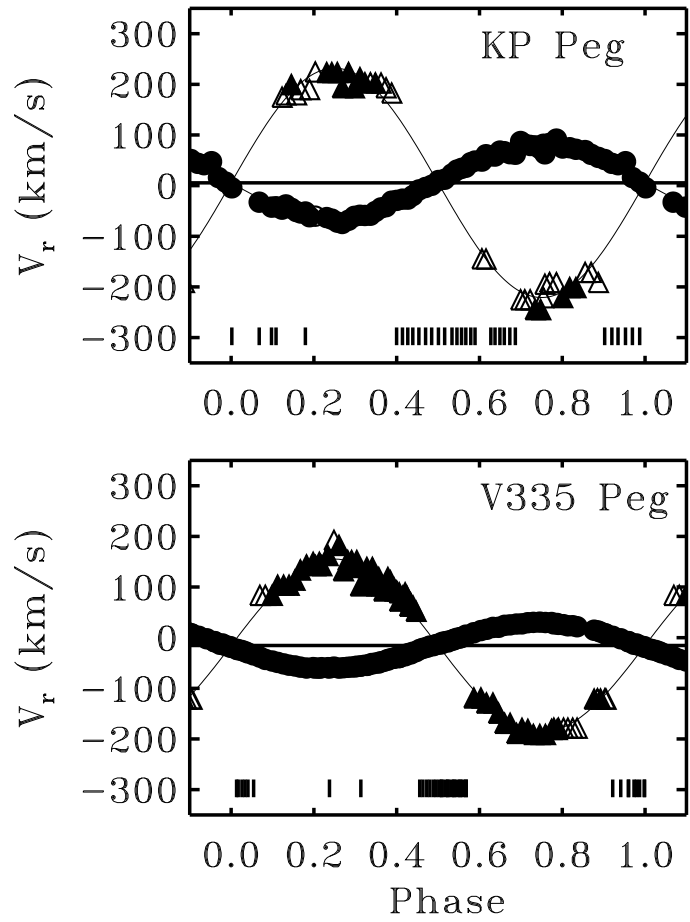}
\caption{
Same as for Figure~1, but with the radial velocity orbits
for the systems KP~Peg and V335~Peg.
Both systems are A-type contact binaries.
}
\label{fig3}
\end{figure}

This paper is structured in a way similar to that of previous
papers, in that most of the data for the observed binaries are in
two tables consisting of the radial-velocity measurements
(Table~\ref{tab1}) and their sine-curve solutions (Table~\ref{tab2}).
The data in Table~\ref{tab2} are organized in the same manner as
in previous papers. In addition to the parameters of spectroscopic orbits,
the table provides information about the relation between
the spectroscopically observed epoch of the primary-eclipse T$_0$
and the recent photometric determinations in the form of the $O-C$
deviations for the number of elapsed periods $E$. It also contains
our new spectral classifications of the program objects.
Section~\ref{sec2} of the paper contains brief summaries of
previous studies for individual systems and comments
on the new data. Figures~\ref{fig1} --
\ref{fig3} show the radial velocity data and solutions.
Figure~\ref{fig4} shows the BF's for all systems;
the functions have been selected from among the
best defined ones around the orbital phase of
0.25 using the photometric system of phases
counted from the deeper eclipse.

\placetable{tab1}

\section{RESULTS FOR INDIVIDUAL SYSTEMS}
\label{sec2}

\subsection{AB And}

Photometric variability of AB~And was discovered by \citet{GP1927}.
\citet{Ooster1930} gave a photometric ephemeris. Twenty years later,
\citet{Ooster1950} reported discovery of the period variation. Since that
time, AB~And became a target of numerous photometric investigations.
On the basis of the photoelectric observations,
\citet{Landolt1969} determined the spectral type of K2V.
He also noted asymmetries in the light curve. The asymmetries have been
explained in \citet{Bell1984} and \citet{Djurasevic2000} by a model with
photospheric spots.
\citet{Demircan1994} suggested that observed period variability may be a
result of the orbital motion in a wide triple system. The third body
should be a white dwarf in such a case.
Str{\"o}mgren photometry presented by \citet{Rucinski1981}
suggested the spectral type of AB~And to be G5.

The first spectroscopic observations of this object were published by
\citet{Struve1950}. AB~And was then classified as a W~UMa star with spectral
type G5. The radial velocities of the components were measured in this
investigation in only 7 spectra, therefore this result was rather preliminary.
The results were: $V_0 = -45$ km~s$^{-1}$, $K_1 = 165$ km~s$^{-1}$
and $K_2 = 265$ km~s$^{-1}$.
An extensive discussion on the system with the combined light and
radial-velocity solution was presented by \citet{Hrivnak1988} who
classified the spectral type of AB~And to be G5V.
Radial velocity curves obtained by \citet{Hrivnak1988} were modified to
include proximity effects. The measured radial velocities gave
following orbital parameters:
$V_0 = -24.6 \pm 0.9$ km~s$^{-1}$
$K_1 = 115.7 \pm 0.7$ km~s$^{-1}$ and
$K_2 = 235.7 \pm 1.5$ km~s$^{-1}$.

For the preliminary moment of eclipse $T_0$, as referred to in
Table~\ref{tab2}, we used the moment of \citet{Pribulla2002}.
Similar to the previous researchers,
we find the system to be a W-type contact binary. Our spectral type,
G8V, is slightly later than most of the previous determinations.
AB~And has relatively red color index $(B-V)=0.925$, which corresponds to
the spectral type K2. Spots in the photospheres of the system components
may be a possible explanation of the spectral type discrepancy. The
broadening functions of both system components are well defined and
radial velocities are measured precisely.
We note a difference in the center-of-mass velocity of the system between
our estimate: $ V_0 = -27.5 \pm 0.7$ km~s$^{-1}$ and the result
presented by \citet{Hrivnak1988}. Also, the amplitude of radial velocity of
the less massive component obtained by \citet{Hrivnak1988} was smaller
than our result, $K_1 = 130.3$  km~s$^{-1}$, possibly
because \citet{Hrivnak1988} used the cross-correlation method which
-- in our experience -- frequently under-estimates the velocity
amplitudes.

The Hipparcos parallax, $8.34 \pm 1.48$ mas (milli-arcsec), gives the
distance of $120 \pm 20$ pc. The observed proper motion is moderately
large \citep{Tycho2},
resulting, for the assumed distance, in tangential velocities of
$V_{RA} = 109$ km~s$^{-1}$ and $V_{dec} = -53$ km~s$^{-1}$ and the
combined spatial velocity of  $V = 74$ km~s$^{-1}$. Because the
observed variability of the period and of
$V_0$ can be interpreted as an influence of
a companion, the parallax may be incorrect; the large error is
consistent with that. The direct, parallax-based estimate
of $M_V = 4.1 \pm 0.4$ only marginally agrees with the one
from the absolute-magnitude calibration \citep{RD97},
$M_V(cal)=5.0$; however, $V_{max}=9.5$ is also
poorly defined, partly because of the spots and partly because
of the small number of calibrated light curves even for this popular system.

The masses of the components are very well defined,
$M_1 \sin^3i = 1.06 M_\odot$ and $M_2 \sin^3i=0.59 M_\odot$,
and are surprisingly large for components of a G8/K2 contact system.
We have seen a very similar situation in AH~Vir \citep{ahvir},
which is also a system consisting of massive but unusually
cool components. It is very likely that the strong magnetic
activity of AB~And and AH~Vir may have something to do with this
anomaly.

\subsection{V402 Aur}

V402~Aur was discovered as a variable by \citet{Oja1991} during an
UBV photometric survey of astrometric standard stars.
The light-curve and ephemeris were published 3 years later \citep{Oja1994}.
The spectral type derived from Henry Draper Extension Charts is F0
\citep{Nesterow1995}. Spectral type in the SIMBAD database is F2.
The spectral type corresponding to $(B-V)=0.40$ derived
from the {\it Tycho-2\/} catalog
\citep{Tycho2} is F3-4. Our new spectral type is F2V.

The mass ratio is small, $q=0.20$. Similar
depths of the eclipses and the well defined broadening functions
strongly suggests that V402~Aur is a contact binary of the W-type
(with the assumed moment of the primary eclipse, as
referred to in Table~\ref{tab2}, of \citet{Pribulla2002}). The
small sum of the masses for an early-F system,
$(M_1 + M_2) \sin^3i = 0.99 \pm 0.03 M_\odot$,
is consistent with the small
photometric amplitude of 0.17 mag., both being due to the
low inclination of the orbit.

\subsection{V445 Cep}

Variability of V445~Cep was discovered by {\it Hipparcos}. The full
amplitude of the {\it Hipparcos} light-curve is only 0.03~mag.
The star has abnormally blue color index for its period
\citep{Rucinski2002b}, which raised a suspicion that
pulsations are the source of the observed photometric variability.
Our radial velocity observations confirm the
binary nature of the system, but with the period equal to twice
the {\it Hipparcos} period. We adjusted the preliminary
{\it Hipparcos} $T_0$ by one quarter forward of the light maximum to
$T_0 = 2,448,500.1146$.

The mass ratio of V445~Cep, $q=0.17 \pm 0.01$, is
small. We also find a very small value of
$(M_1 + M_2) \sin^3 i = 0.134 \pm 0.006 M_\odot$.
This result, together with small photometric amplitude, suggest a
small inclination angle of the orbit. The
system seems to be a contact binary, but small amplitudes of radial
velocities and photometric variability make it difficult to derive
the orbital parameters.

The results of low-resolution
spectroscopic observations were presented by
\citet{Grenier1999}. The star was classified as A2V. The radial
velocity of the whole system $V_r=38.6 \pm 9.6$ km~s$^{-1}$
is in a good agreement with our $V_0 = 40.69 \pm 0.95$
km~s$^{-1}$.

The color index $(B-V)=0.123$ was found in the {\it Tycho-2\/}
catalog \citep{Tycho2}. This corresponds to a spectral type A4.
Our spectral type is A2V.
The system is bright, $V_{max}=6.82$, which together with the
{\it Hipparcos} parallax gives, $M_V=1.58 \pm 0.13$. Thus, this
is one of the best determined luminosities for an A spectral-type
contact binary.

\subsection{V2082 Cyg}

This star was listed as a variable candidate by \citet{Hoffleit1979}.
It's variability was confirmed by {\it Hipparcos}.
The light-curve from {\it Hipparcos} has a small amplitude
of only 0.05~mag and similar depths of the eclipses. We used the
{\it Hipparcos} data for the preliminary $T_0$.

The spectral type of V2082~Cyg in the SIMBAD database is F0, while
our spectral type is F2V. $(B-V)=0.313$ from
{\it Tycho-2\/}  catalog \citep{Tycho2}
corresponds to the spectral type F1.
V2082~Cyg is most probably an A-type contact binary, although
the secondary component is faint (the relative luminosity
from the broadening function, $L_2 = 0.10 \pm 0.02$),
so that we cannot exclude
a semi-detached configuration. The system must be viewed
at a very low inclination angle which would explain the
small photometric and radial velocity amplitudes, leading too
poorly resolved broadening functions, with the partly merged
signatures of both components (see Fig.~\ref{fig4}).

The absolute magnitude based on the {\it Hipparcos} parallax
and the adopted $V_{max}=6.64$ is $M_V=1.85 \pm 0.12$. It agrees
well with the RD97 calibration, $M_V(cal)=1.71$. The proper
motion, especially in declination, is large \citep{Tycho2},
resulting, for the assumed distance, in the
combined spatial velocity of $V = 54$ km~s$^{-1}$.

The radial velocity of the system was previously measured
at low resolution by \citet{Shajn1951} on three
occasions. The author noticed that the radial
velocity of this star is variable.

\subsection{BX Dra}

BV 228 (BX Dra) was discovered to be
a variable star by \citet{Strohmeier1958}.
\citet{Strohmeier1965} classified the star as an RR~Lyrae type
variable. Some doubt to this classification and hints about the binary
nature of the star based both on spectroscopy and photometry were
presented by \citet{Smith1990}. The author, in a
photometric survey of variable stars,
found a new period and changed the classification to
an elliptical variable.
Independently, \citet{Agerer1995} presented a light-curve and
classified BX Dra as an $\beta$~Lyr type variable.
Nevertheless \citet{Solano1997} still regarded it as a RR~Lyr variable,
although they have found some other stars to be incorrectly classified
as RR~Lyr variables in the General Catalogue of Variable Stars
\citep{Kholopov1985}.
SIMBAD still lists this star as a variable star of RR~Lyr type.
The spectral type of BX~Dra in the SIMBAD database is A3.

\begin{figure}
\figurenum{4}
\epsscale{0.7}
\plotone{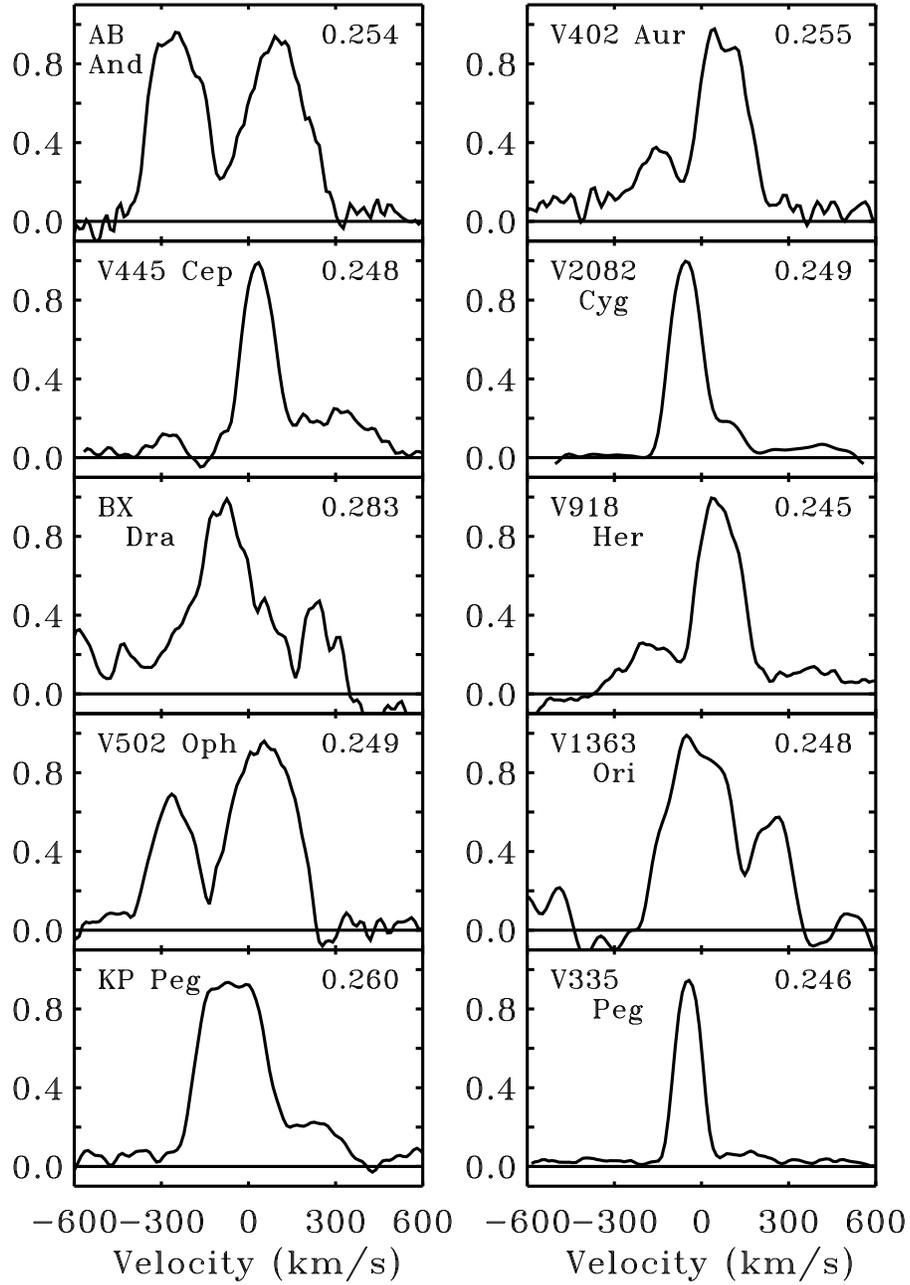}
\vspace{10mm}
\figcaption{
The broadening functions (BF's) for all binary systems of this
group; all for orbital phases around 0.25, as in similar figures
in the previous papers.
}
\label{fig4}
\end{figure}

We found that the variable is definitely a contact binary of the
A-type. The color index $(B-V)=0.352$ from {\it Tycho-2\/} catalog
\citep{Tycho2} corresponds to the spectral type of F2.
Our spectral type is F0IV-V.
The star is one of the faintest in our sample
($V_{max}=10.5$) and the broadening functions
are noisy and show undesirable baseline slopes
(see Fig.~\ref{fig4}). However,
because the radial velocity amplitudes are
relatively large, this object may deserve further photometric
and spectroscopic investigations.

The radial velocity has been  measured by \citet{Layden1994} at low
resolution, $V_r=75 \pm 30$ km~s$^{-1}$. The
radial velocity was also measured again by \citet{Solano1997},
and the result $ V_r=-24 \pm 3$ km~s$^{-1}$,
is in agreement with our result
$ V_r = -26.11 \pm 3.43$ km~s$^{-1}$ within the respective errors.
The moment of the primary eclipse referred to in Table~\ref{tab2}
was taken from the {\it Hipparcos} catalog.

\subsection{V918 Her}

Variability of this star was discovered by {\it Hipparcos}.
The spectral type of this star in the SIMBAD database is A2.
\citet{Grenier1999} classified its spectrum as A5V.
Our spectral classification of V918~Her is A7V.
The $(B-V)$ index from the
{\it Tycho-2\/} catalog \citep{Tycho2}
is 0.249 and corresponds to a spectral
type A8/9V, which may indicate some reddening.

The radial velocity of the system measured by \citet{Grenier1999}
$V_r=-33.9 \pm 7.2$ km~s$^{-1}$ is different from
our result of $V_r=-25.72 \pm 0.74$  km~s$^{-1}$.

We find the object to be an A-type contact binary. Otherwise the system
is rather inconspicuous, but is bright, $V_{max}=7.30$, and thus
was included in the magnitude-limited sample of
\citet{Rucinski2002b}.
We have adopted $T_0$ from the {\it Hipparcos} catalog.

\subsection{V502 Oph}

V502 Oph was discovered to be an eclipsing binary by
\citet{Hoffmeister1935}. The first ephemeris based on visual observations
was published by \citet{Lause1937}.
Over the years, the star has been the subject of numerous investigations.
The light curve of the system is not stable and the orbital period was
found to undergo a change in the years 1955--1966 \citep{Binnendijk1969}.
The period is successively decreasing, but the rate of the change 
observed in the years 1989 -- 2003 has not been 
constant \citep{Kreiner2003}.
For Table~\ref{tab2}, we adopted $T_0$ from the {\it Hipparcos} catalog.
We found however, that the orbital period from the {\it Hipparcos}
catalog definitely does not fit our
spectral data. We established that the period
which fits our data best is 0.453390~days and this period was used
for calculating the orbital elements in Table~\ref{tab2}.

Observations of V502~Oph with the VLA revealed that it is a binary radio
source \citep{Hughes1984}. Since W~UMa-type systems usually show low
radio activity \citep{Rucinski1995},
this may suggest the existence of an optically
undetected companion to the eclipsing binary system \citep{Hughes1984}.
The presence of a late-type tertiary component was in fact noticed in the
spectrum of V502~Oph by \citet{Hendry1998}.

The pioneering investigation on the radial velocity orbit
of this W~UMa type variable was done by Gratton
(as described in \citet{Struve1948}).
The derived orbit elements based on these
old observations, $V_r = -37$ km~s$^{-1}$,
$K_1 = 95$ km~s$^{-1}$, $K_2 = 235$ km~s$^{-1}$,
are surprisingly close the values presented in Table~\ref{tab2}.
Radial velocity measurements presented later
by \citet{Struve1959} generally supported earlier results;
the $-13$~km~s$^{-1}$ shift in the mass-center velocity
was considered insignificant in view of the low
accuracy of the measurements.
The spectra of the components were classified as G1V for the primary
and F9V for the secondary component, reflecting the fact that this
is a W-type contact system with a slightly hotter secondary.

The spectral type in the SIMBAD database is "G2V+...".
The spectral type of V502~Oph based on the {\it Tycho-2\/}  color index
\citep{Tycho2}, $(B-V)=0.615$, is G1, which is close to G0V found by us.

\subsection{V1363 Ori}

The variability of V1363~Ori was discovered by {\it Hipparcos}.
The spectral type derived from Henry Draper Extension Charts is F5
\citep{Nesterow1995}, while in the SIMBAD database it is F8.
The $(B-V)=0.56$ color index derived from the {\it Tycho-2\/}
catalog \citep{Tycho2} corresponds to a spectral type of F9.
For technical reasons, we were not able to obtain a good
classification spectrum of the star; we can only confine
it to the early to mid-F spectral type.

The light-curve of this variable presented by
\citet{Gomez-Forrellad1999} shows the O'Connell effect. This source
was used for the initial $T_0$.

V1363~Ori is an A-type contact binary at the faint end of our
magnitude accessibility, so that the broadening functions are rather
poor. The {\it Hipparcos} parallax is poorly determined,
$9.47 \pm 2.36$, so that the binary may have a spectroscopically
undetectable companion.

\subsection{KP Peg}

This star was listed as a suspected variable by \citet{Hopmann1948}.
\citet{Walker1987} confirmed that the component A of this visual binary system
(separation 3.5 arcsec, magnitude difference 1.6)
is a variable of Beta Lyrae type. He also gave its ephemeris and presented
the light-curve \citep{Walker1988}.
\citet{Abt1985}, in one of his spectral classification papers of binaries
from the \citet{Aitken1932} catalog (ADS), classified it as an A2V star.
We confirm this spectral type. It is also consistent with the
$(B-V)=0.060$ color index from the {\it Tycho-2\/} catalog \citep{Tycho2}.
Due to the early spectral type and the weak spectral lines in our
standard spectral window, the broadening functions are rather
poorly defined, with the component signatures partly merged
in the broadening functions.

The {\it Hipparcos} parallax, $4.37 \pm 1.67$ mas, is
poorly determined, probably because of the presence of the third body
in the system. With the corrected $V_{max}=7.07$ \citep{Rucinski2002b},
the absolute magnitude of the binary is one of the brightest in our
program, $M_V=+0.3 \pm 0.8$.
The adopted $T_0$ is from the {\it Hipparcos} catalog.

\subsection{V335 Peg}

Variability of V335~Peg was discovered by {\it Hipparcos}.
The light-curve from {\it Hipparcos} has a small amplitude of 0.05~mag.
It is most probably an A-type contact binary, although
the broadening function of the
secondary component is very weak (the relative luminosity
determined from the broadening function is only
$L_2 = 0.05 \pm 0.01$) and is difficult to measure, contrary to that
for the primary component whose velocity could be measured
very precisely. At this point we cannot exclude
a semi-detached configuration.

When we phased the observations,
we noticed a small -- about $3.5$~km~s$^{-1}$,
systematic shift in radial velocities of the primary component between
seasons 2001 and 2002. We have found that a
correction to the period derived from {\it Hipparcos} light-curve may
eliminate this discrepancy. The orbital solution
presented in Table~\ref{tab2} has been obtained with the new period,
 P=0.81072~days, but with the same $T_0$. We note that this period implies
a large $T_0$ shift of $O-C = -0.159$ days.
The orbital parameters obtained for the {\it Hipparcos}
period: P=0.810746~days, $V_0 = -15.85 \pm 0.47$~km~s$^{-1}$,
K$_1 = 45.52 \pm 0.33$~km~s$^{-1}$,
K$_2 = 169.58 \pm 1.77$~km~s$^{-1}$, result in a
smaller $O-C = -0.036$ days.
In the absence of any photometric data during the
elapsed 4724 orbital cycles, we have not been able to decide if the
original period was incorrect or was variable, or there was
a radial-velocity shift caused by a third body in the system with
a possible miscount in the number of orbital cycles.

V335~Peg is one of the brightest short-period binaries
in the sky \citep{Rucinski2002b},
$V_{max}=7.24$, and is relatively nearby with the parallax
$16.26 \pm 0.86$ mas, resulting in $M_V=3.30 \pm 0.12$. The color index
$(B-V)=0.439$ corresponds to the spectral type of F5, which is
the same as found in our spectra, F5V. This does not agree well
with the RD97 calibration, $M_V(cal)=1.85$. The reason for this
discrepancy is not clear. We note that
V335~Peg is a source of X-ray radiation and is listed in ROSAT Bright
Survey \citep{Schwope2000}.
The proper motions of V335~Peg are large
in both coordinates \citep{Tycho2}, which --
when coupled with the moderate distance of 62 pc --
results in a combined spatial velocity of $V = 58$ km~s$^{-1}$.

\section{SUMMARY}

This paper presents spectral classifications, radial velocity data and
circular orbital solutions for the eighth group of ten close binary
systems observed at the David Dunlap Observatory.
All systems are double-lined (SB2) contact binaries.
Half of the systems (V445~Cep, V2082~Cyg, V918~Her, V1363~Ori, V335~Peg)
were discovered photometrically by the {\it Hipparcos} mission and
two are well known, frequently observed contact systems
(AB~And, V502~Oph) which had been previously observed spectroscopically,
but for which our broadening function method permitted improvement
of the orbital elements. We spectroscopically detected
very weak companions of V2082~Cyg, KP~Peg and especially
V335~Peg. We note that V445~Cep, V2082~Cyg, V918~Her,
KP~Peg and V335~Peg are bright binaries with the observed
$V_{max}<7.5$. They were previously considered in the rigorously
selected, magnitude-limited sample of \citet{Rucinski2002b}.

\acknowledgements

This study has been done while W. Pych held the NATO
Post-Doctoral Fellowship administered by the
Natural Sciences and Engineering Council of Canada (NSERC); he also
acknowledges the support from the Polish Grant KBN 2~P03D~029~23.
The NSERC supported research of S. M. Rucinski and of 
R. M. Blake, through a research grant to T. Bolton.
W. Ogloza, G. Stachowski and K. Gazeas acknowledge the 
travel and subsistence support from the 
NATO collaborative linkage grant PST.CLG.978810 as well
as the Polish KBN grant 2-P03D-006-22.

The research has made use of the SIMBAD database, operated at the CDS,
Strasbourg, France and accessible through the Canadian
Astronomy Data Centre, which is operated by the Herzberg Institute of
Astrophysics, National Research Council of Canada.

\clearpage                    

\begin{table}                 
\dummytable \label{tab1}      
\end{table}

\begin{table}                 
\dummytable \label{tab2}      
\end{table}


\begin{thebibliography}{}
\bibitem[Abt(1985)]{Abt1985}			      	
    Abt, H. A. 1985, \apjs, 59, 95
\bibitem[Agerer \& Dahm(1995)]{Agerer1995}		
    Agerer, F., \& Dahm, M. 1995, IBVS, 4266, 1
\bibitem[Aitken(1932)]{Aitken1932}				
    Aitken, R.G. 1932, New General Catalogue of Double Stars within
    120\degr of the North Pole
    (Carnegie Inst. of Washington Pub., No. 417)(ADS).
\bibitem[Bell, Hilditch, \& King(1984)]{Bell1984}			
    Bell, S. A., Hilditch, R. W., \& King, D. J.
    1984, \mnras, 208, 123
\bibitem[Bessell(1979)]{Bessell1979}			
    Bessell, M. S. 1979, \pasp, 91, 589
\bibitem[Binnendijk(1969)]{Binnendijk1969}			
    Binnendijk, L. 1969, \aj, 74, 218
\bibitem[Demircan et~al.(1994)]{Demircan1994}		
    Demircan, O., Derman, E., Akalin, A., Selam, S., \& Muyesseroglu, Z.
    1994, \mnras, 267, 19
\bibitem[Djurasevic, Rovithis-Livaniou, \& Rovithis(2000)]{Djurasevic2000}	
    Djurasevic, G., Rovithis-Livaniou, H., \& Rovithis, P.
    2000, \aap, 364, 543
\bibitem[Grenier et~al.(1999)]{Grenier1999}		
    Grenier, S., Baylac, M.-O., Rolland, L., Burnage, R., Arenou, F., Briot,
    D., Delmas, F., Duflot, M., Genty, V., Gómez, A. E., Halbwachs, J.-L.,
    Marouard, M., Oblak, E., \& Sellier, A.
    1999, \aaps, 137, 451
\bibitem[Gomez-Forrellad et~al.(1999)]{Gomez-Forrellad1999}	
    Gomez-Forrellad, J. M., Garcia-Melendo, E., Guarro-Flo, J.,
    Nomen-Torres, J., \& Vidal-Sainz, J.
    1999, IBVS, 4702
\bibitem[Guthnik \& Prager(1927)]{GP1927}       	
    Guthnick, P., \& Prager, R.
    1927, Kleinere Ver{\"o}ff. Berlin-Babelsberg, No. 4, 16
\bibitem[Hendry \& Mochnacki(1998)]{Hendry1998}
    Hendry, P. D., Mochnacki, S. W.
    1998, \apj, 504, 978
\bibitem[Hoffleit(1979)]{Hoffleit1979}			
    Hoffleit, D. 1979, BICDS, 17, 24
\bibitem[Hoffmeister(1935)]{Hoffmeister1935}		
    Hoffmeister, C.
    1935, Astron.\ Nachr., 255, 403
\bibitem[H{\o}g et~al.(2000)]{Tycho2}
    H{\o}g, E., Fabricius, C., Makarov, V. V., Urban, S., Corbin, T.,
    Wycoff, G., Bastian, U., \\
    Schwekendiek, P., \& Wicenec, A.
    2000, \aap, 355L, 27
\bibitem[Hopmann(1948)]{Hopmann1948}			
    Hopmann, J.
    1948, Z. Astrophysics, 24, 263
\bibitem[Hrivnak(1988)]{Hrivnak1988}			
    Hrivnak, B. J.
    1988, \apj, 335, 319
\bibitem[Hughes \& McLean(1984)]{Hughes1984}
    Hughes, V. A., McLean, B. J.
    1984, \apj, 278, 716
\bibitem[Kholopov et~al.(1985-1988)]{Kholopov1985}	
    Kholopov, P. N., et~al. 1985-1988, General Catalogue of Variable
    Stars (4th ed.; Moscow: Nauka)
\bibitem[Kreiner(2003)]{Kreiner2003}
    Kreiner, J. M. 2003, private communication
\bibitem[Landolt(1969)]{Landolt1969}			
    Landolt, A. U.
    1969, \aj, 74, 1078
\bibitem[Lause(1937)]{Lause1937}				
    Lause, F. 1937, Astron. Nachr., 264, 106
\bibitem[Layden(1994)]{Layden1994}				
    Layden, A. C. 1994, \aj, 108, 1016
\bibitem[Lu \& Rucinski(1993)]{ahvir}			
    Lu, W., \& Rucinski, S. M.
    1993, \aj, 106, 361
\bibitem[Lu \& Rucinski(1999)]{ddo1}			
    Lu, W., \& Rucinski, S. M.
    1999, \aj, 118, 515 (Paper I)
\bibitem[Lu, Rucinski, \& Ogloza(2001)]{ddo4}			
    Lu, W., Rucinski, S. M., \& Ogloza, W.
    2001, \aj, 122, 402 (Paper IV)
\bibitem[Nesterow et~al.(1995)]{Nesterow1995}		
    Nesterov, V. V., Kuzmin, A. V., Ashimbaeva, N. T.,
    Volchkov, A. A., Röser, S., \& Bastian, U.
    1995, \aaps, 110, 367
\bibitem[Oja(1991)]{Oja1991}					
    Oja, T. 1991, \aaps, 89, 415
\bibitem[Oja(1994)]{Oja1994}					
    Oja, T. 1994, IBVS, 4000, 1
\bibitem[Oosterhoff(1930)]{Ooster1930}			
    Oosterhoff, P. Th.
    1930, \bain, 5, 151
\bibitem[Oosterhoff(1950)]{Ooster1950}			
    Oosterhoff, P. Th.
    1950, \bain, 11, 217
\bibitem[Pribulla et~al.(2002)]{Pribulla2002}
    Pribulla, T., Vanko, M., Parimucha, S., Chochol, D.
    2002, IBVS, 5341, 1
\bibitem[Pych(2003)]{Pych2003}
    Pych, W., 2003,  \pasp, submitted (astro-ph/0311290)
\bibitem[Rucinski(1995)]{Rucinski1995}		
     Rucinski, S. M
     1995, \aj, 109, 2690
\bibitem[Rucinski(2002a)]{ddo7}			
     Rucinski, S. M
     2002a, \aj, 124, 1746 (Paper VII)
\bibitem[Rucinski(2002b)]{Rucinski2002b}	
     Rucinski, S. M.
     2002b, \pasp, 114, 1124
\bibitem[Rucinski et~al.(2003)]{ddo8}			
     Rucinski, S. M., Capobianco, C. C., Lu, W.,
     DeBond, H.,  Thomson, J. R., Mochnacki, S. W., Blake, R. M.,
     Ogloza, W., Stachowski, G., \& Rogoziecki, P.
     2003, \aj, 125, 3258 (Paper VIII)
\bibitem[Rucinski \& Duerbeck(1997)]{RD97}	
    Rucinski, S. M., \& Duerbeck, H.W. 1997,
    \pasp, 109, 1340  (RD97)
\bibitem[Rucinski \& Kaluzny(1981)]{Rucinski1981}
    Rucinski, S. M., \& Kaluzny, J.
    1981, AcA, 31, 409
\bibitem[Rucinski \& Lu(1999)]{ddo2}			
    Rucinski, S. M., \& Lu, W.
    1999, \aj, 118, 2451 (Paper II)
\bibitem[Rucinski et~al.(2002)]{ddo6}			
     Rucinski, S. M., Lu, W., Capobianco, C. C.,
     Mochnacki, S. W., Blake, R. M., Thomson, J. R.,
     Ogloza, W., \& Stachowski, G.
     2002, \aj, 124, 1738 (Paper VI)
\bibitem[Rucinski, Lu, \& Mochnacki(2000)]{ddo3}		
     Rucinski, S. M., Lu, W., \& Mochnacki, S. W.
     2000, \aj, 120, 1133 (Paper III)
\bibitem[Rucinski et~al.(2001)]{ddo5}				
     Rucinski, S. M., Lu, W., Mochnacki, S. W.,
     Ogloza, W., \& Stachowski, G.
     2001, \aj, 122, 1974 (Paper V)
\bibitem[Solano et~al.(1997)]{Solano1997}			
    Solano, E., Garrido, R., Fernley, J., \& Barnes, T. G.
    1997, \aaps, 125, 321
\bibitem[Smith(1990)]{Smith1990}				
    Smith, H. A. 1990, \pasp, 102, 124
\bibitem[Schwope et~al.(2000)]{Schwope2000}		
    Schwope, A., Hasinger, G., Lehmann, I., Schwarz, R., Brunner, H.,
    Neizvestny, S., Ugryumov, A., Balega, Yu., Tr{\"u}mper, J., \& Voges, W.
    2000, Astr. Nachr., 321, 1
\bibitem[Strohmeier(1958)]{Strohmeier1958}		
    Strohmeier, W.
    1958, Kl. Veröffent.\ der Sternwarte Bamberg, No. 24
\bibitem[Strohmeier et~al.(1965)]{Strohmeier1965}	
    Strohmeier, W. et~al.,
    1965, Veröffent. der Sternwarte Bamberg, Bd., V, No. 18
\bibitem[Struve \& Gratton(1948)]{Struve1948}		
    Struve, O., Gratton, L.
    1948, \apj, 108, 497
\bibitem[Struve et~al.(1950)]{Struve1950}		
    Struve, O., Horak, H. G., Canavaggia, R., Kourganoff, V.,
    \& Colacevich, A.
    1950, \apj, 111, 658
\bibitem[Struve \& Zebergs(1959)]{Struve1959}		
    Struve, O., Zebergs, V.
    1959, \apj, 130, 789
\bibitem[Shajn(1951)]{Shajn1951}			
    Shajn, G. A.
    1951, Bull.\ Crimean Astrophys.\ Obs., 7, 124
\bibitem[Walker(1987)]{Walker1987}			
    Walker, R. L.
    1987, \baas, 19, 1085
\bibitem[Walker(1988)]{Walker1988}			
    Walker, R. L.
    1988, IBVS, 19, 1085
\end{thebibliography}
\end{document}